\documentclass[manuscript,sigconf]{acmart}
\AtBeginDocument{%
  \providecommand\BibTeX{{%
    \normalfont B\kern-0.5em{\scshape i\kern-0.25em b}\kern-0.8em\TeX}}}

\setcopyright{acmcopyright}
\copyrightyear{2018}
\acmYear{2018}
\acmDOI{XXXXXXX.XXXXXXX}

\usepackage{caption}
\usepackage{subcaption}

\begin{document}

\title{QR TPM in Programmable Low-Power Devices}

\author{Luís Fiolhais and Leonel Sousa}
\email{luis.azenhas.fiolhais@tecnico.ulisboa.pt,  las@inesc-id.pt}
\orcid{1234-5678-9012}
\authornotemark[1]
\affiliation{%
  \institution{INESC-ID, Instituto Superior Técnico, Universidade de Lisboa}
  \streetaddress{Rua Alves Redol, 9}
  \city{Lisboa}
  \country{Portugal}
  \postcode{1000-029}
}

\begin{abstract}
Trusted Platform Modules (TPMs), which serve as the root of trust in secure systems, are secure crypto-processors that carry out cryptographic primitives. Should large-scale quantum computing become a reality, the cryptographic primitives adopted in the TPM 2.0 standard will no longer be secure. Thus, the design of TPMs that provide Quantum Resistant (QR) primitives is of utmost importance, in particular with the restrictions imposed by embedded systems. In this paper, we investigate the deployment of QR primitives and protocols in the standard TPM 2.0.  Cryptographic algorithms that are already in the NIST QR cryptography standardization process, as well as an Oblivious Transfer (OT), a fundamental cryptographic primitive, are the QR cryptographic schemes selected to extend TPM 2.0. 
In particular, the Kyber algorithm for key encapsulation, the Dilithium  algorithm for digital signature, and a 3-round Random Oblivious Transfer (ROT) protocol, supporting protocols such as Multi-Party Computation and Private Set Intersection (PSI). The QR extended TPM 2.0 is  implemented in ARM and RISC-V embedded processors, its computational requirements are
analysed and experimentally evaluated in comparison to the standard TPM. It is shown that Kyber and Dilithium are faster at creating keys than RSA, due to the key size and secure random sampling required in RSA, while  they meet the same
performance level as ECC. For digital signatures, both in signature creation and verification, Dilithium is on par with RSA and ECC. The ROT protocol shows decent performance and its
support required small modifications to the TPM. 
This paper also shows that it would be possible to backport the required code to already available TPMs to ensure that current TPMs remain secure against quantum adversaries.

\end{abstract}



\keywords{Trusted Platform Module, Cryptography Quantum-Resistant, Embedded Systems}



\maketitle

\section{Introduction}

The Trusted Platform Module (TPM) provides hardware-based support to assure the security of computers and embedded devices. TPM provides various security functions, such as encryption, secure key storage, and authentication, to ensure the integrity of a system and protect sensitive information. Fundamental functions and features of TPM are: secure boot; authentication of themselves to other devices, by storing cryptographic keys and certificates securely; data encryption, to perform encryption and decryption operations securely; and remote attestation, for remote verification and attestation.

The current TPM standard relies on number-theoretic cryptography~\cite{tpm}. However, should quantum computers become available, these types of cryptographic primitives, such as Rivest-Shamir-Adleman (RSA) and Elliptic Curve Cryptography (ECC), will no longer be secure~\cite{Shor97}. Thus, the National Institute of Standards and Technology (NIST) initiated, a few years ago, a process to solicit, evaluate, and standardize quantum-resistant (QR) public-key cryptographic algorithms~\cite{nist-pq}. 

The NIST has already identified four candidate QR cryptographic algorithms for standardization. Among these, the two primary algorithms, members of the Cryptographic Suite for Algebraic Lattices (CRYSTALS), are the CRYSTALS-Kyber~\cite{bos18} for providing key encapsulation mechanisms, and the CRYSTALS-Dilithium~\cite{ducas19} for digital signatures. These algorithms are based on the Module Ring Learning With Errors (MLWE) problem and the Ring Learning With Errors (RLWE) problem. It is conjectured that these problems are hard for both classical and quantum computers~\cite{EC:LyuPeiReg10}. Supported also on the RLWE problem, Oblivious Transfer (OT) and Random Oblivious Transfer (ROT) primitives have been proposed~\cite{Branco21}. These primitives  are important for supporting  protocols such as Multi-Party Computation and Private Set Intersection (PSI), which are used in applications like contact discovery, remote diagnosis, and contact tracing.

Although components complying with  the TPM 2.0 standard are stable and widely available, QR primitives have to be introduced in it  for extending the actual capabilities and assure the security of a computing platform. Recently, research projects, such as the FutureTPM~\cite{fiolhais2020software}, have advanced in this topic, and non-Quantum Resistant TPMs have been extended through software emulation~\cite{fiolhais2020software}. It was concluded that Kyber and Dilithium can efficiently replace most of the functionality provided by ECC and RSA. However, there are others, such as the OT protocol, that have not still been evaluated in the scope of the current TPM. Moreover, now that QR cryptographic primitives are already in the standardization phase, hardware has to be developed for TPM to keep pace with QR cryptography. 

The change of paradigm to QR raises a new set of challenges, in particular on embedded devices, including ones related to performance and memory requirements, but also related to side-channel security. In this paper, we analyse the computational requirements of an updated QR TPM based on low-end range processors. To experimentally validate and evaluate a QR TPM, we have chosen an in-order double-issue RISC-V U74 processor and the in-order double-issue ARM Cortex-A7. The adoption of programmable processors is the right choice in this first stage of development, not only allowing adjustments and parametrization during the standardization phase but also assessing  methods to face the challenges of side-channel attacks. As far as we know, this is the first time this type of research, with significant practical interest, is performed, being a step forward   
for launching secure and efficient QR TPM devices in the near future. Moreover, our approach demonstrates that
it would be possible to backport the required code to already available TPMs in order to ensure that old TPMs
remain secure against quantum adversaries.

The organization of the paper is as follows. In Section~\ref{sec:background} we provide the background  information required to understand the material in this paper and briefly discuss the state-of-the-art. Section~\ref{sec:requirements} analyses the memory and computational requirements of the Kyber, Dilithium, and the ROT algorithms. The experimental results presented in Section~\ref{sec:experimental} were obtained by implementing the algorithms and protocol on ARM and RISC-V based embedded systems. Section~\ref{sec:conclusions} draws the conclusions.

\section{Background and State of the Art}
\label{sec:background}
The main objective of this paper is to analyse the computational requirements of QR TPMs, and to experimentally evaluate in practice the feasibility of these new types of secure TPMs by using low-end range programmable microprocessors. In this section, we will discuss the current TPM standard and the QR algorithms and primitives added to this standard classic TPM. 

\subsection{Trusted Platform Module}

The TPM is a specialized hardware component designed to provide security-related functions for computing systems
The standard 
TPM~2.0 library, from the Trusted Computing Group (TCG)~\cite{TCG2016}, builds upon the foundation of TPM 1.2 with enhanced capabilities. In relation to TPM 1.2,  TPM 2.0 provides enhanced cryptography, for example, ECC and more advanced hash functions, allows for asymmetric key encryption directly within the TPM, provides multiple hierarchies of keys, and facilitates remote management and configuration of the TPM.  Although TPM 2.0 is designed to meet the evolving security requirements of current  computing platforms, such as laptops, servers, embedded systems, and Internet of Things (IoT) devices, they are not quantum resistant. With that purpose, QR cryptographic primitives and algorithms have to be implemented in hardware and integrated into the TPM.

\subsection{Kyber and Dilithium algorithms}
Kyber~\cite{bos18} and Dilitium~\cite{ducas19} are QR cryptographic schemes based on lattice-based cryptography, which relies on the hardness of mathematical problems related to lattices. They have been  identified for NIST as QR cryptographic algorithms for standardization. 

Both algorithms require the implementation of sampling, number theoretic transform (NTT) to speedup arithmetic, and encode/decode functions. The Kyber algorithm provides secure key encapsulation mechanisms, for exchanging encryption keys between two parties. It aims to achieve a balance between security and efficiency, having reasonable computational requirements. A standalone hardware design of the Kyber algorithm, which computes key-generation, encapsulation (encryption), and decapsulation (decryption and reencryption), can be found in~\cite{Xing_Li21}.  

The Dilithium digital signature scheme has been designed for digital signature generation and verification, with the aim of being secure and  efficient in terms of computational and memory requirements. It allows multiple security levels through parametrization. The Dilithium family includes different parameter sets that offer varying levels of security and performance, which can be chosen according to the security requirements and available resources. An implementation of Dilithium, for a set of parameters, targeting Field Programmable Gate Arrays (FPGAs) is proposed in~\cite{Beckwith21}.

\subsection{ROT Protocol}

Random Oblivious Transfer (ROT) is a cryptographic protocol that allows one party (the sender) to send a set of messages to another party (the receiver), who can then select and receive a single message from the set without revealing which message was chosen to the sender. This protocol provides a form of private communication where the sender remains oblivious to the choice made by the receiver, and the receiver remains oblivious to the content of the unchosen messages.

The concept of Oblivious Transfer (OT) was introduced in the field of cryptography to address scenarios where one party needs to transfer information to another party without revealing unnecessary details. ROT extends this concept by introducing an element of randomness, making it more suitable for certain applications. There are different variations of ROTs, including:

\begin{itemize}
    \item 1-out-of-2 ROT (1-2 ROT): In this type of ROT, the sender has two messages, and the receiver chooses one of them without revealing the choice. This is similar to flipping a coin where the sender does not know the outcome, and the receiver only gets the chosen side.

    \item k-out-of-N ROT (k-N ROT): This is a more general version where the sender has N messages, and the receiver can choose k messages without revealing the choices or the content of the unchosen messages.
\end{itemize} 

ROT has applications in secure multiparty computation, private database queries, and cryptographic protocols where privacy and confidentiality are essential. It's a fundamental building block in constructing secure protocols that allow parties to interact without revealing sensitive information. As 
a practical example consider that the TPM is being used as a secure wallet and that the user wants to
perform a payment anonymously. The usage of ROT within the TPM would be advantageous as the contents of the
secure wallet would never have to leave the TPM when a transaction is processed, and the transfer can be
anonymous~\cite{chiu2023tpmwallet}. Moreover, consider the example of
anonymous remote attestation. An enterprise wants to authenticate that a laptop belongs to its network such
that it can give
the laptop access to some confidential internal documents. Once again, the TPM, which already features functionality
for attestation~\cite{TCG2016}, can be used to anonymize the attestation procedure~\cite{brickell2004direct}.

\subsection{Target Hardware: Microprocessors selection}
\label{harware}

Previous work in this topic has explored the architectural, performance, and memory requirements of QR algorithms in a TPM~\cite{fiolhais2020software}. However, the results presented therein used a laptop-class processor, with an out-of-order backend, to emulate the TPM hardware, thus the objective was not to evaluate performance in a real world scenario. Even though it would be beneficial for a TPM to have an out-of-order processor, this is not a realistic goal. TPMs are designed for low power and low price such that they can be used in any type of computing system. The goal of the TPM is to lower the barrier of entry for any computing system to have strong security guarantees. Therefore, this paper aims to perform a performance comparison using cheaper and low-power devices. In these experiments, two low-power processors are used: an ARM Cortex-A7 at 900 MHz and a RISC-V U74 core at 1.2 GHz. Both of these processors were selected as they fit the power/performance/area requirements
found in hardware TPMs. 

The ARM Cortex-A7 is a 32-bit dual-issue in-order processor with support for NEON
instructions. It has 8 pipeline stages and can issue up to two instructions per cycle under
certain conditions. The A7 core used in the experiments is part of the BCM2836 System-on-Chip
(SoC), which features a cluster of four A7 cores. It is unclear what cache hierarchy is
being used in the SoC~\cite{arm-cortex-a7}. 

The RISC-V
U74 is a 64-bit RV64IMAFDC double-issue in-order processor with no vector unit. It has
8 pipeline stages and can issue up to two instructions per cycle under certain
conditions. The U74 core used in the experiments is part of the FU740 SoC, which features
a cluster of four U74 cores~\cite{sifive-fu740}. This SoC has a private 32 KiB 4-way
I\$ and 32 KiB 8-way D\$ for each U74 core, and a 2 MiB 16-way L2 cache shared between all
cores in the cluster.

\section{TPM Computational Requirements}
\label{sec:requirements}

\begin{figure}[htb]
  \centering
  \includegraphics[width=0.25\textwidth]{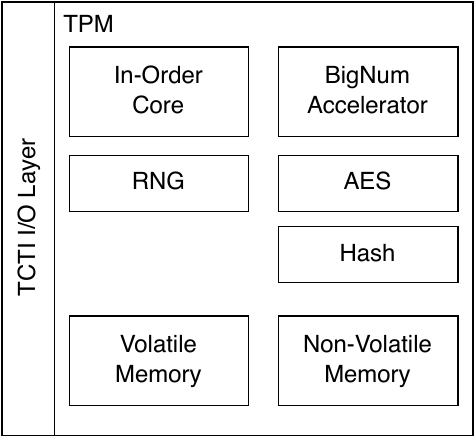}
  \caption{Example Number-Theoretic TPM Architecture.}
  \label{fig:orig-arch}
\end{figure}

Figure~\ref{fig:orig-arch} shows the basic architecture of the TPM. The base architecture is composed of a cryptographic processor wherein a secure Random Number Generator (RNG), RSA and ECC cryptographic primitives, and a hashing engine are available; a small non-volatile memory module (64 kB) to store TPM's
state; and a volatile memory to keep short-lived data. 
The TPM is both a passive and active
agent in a system. It provides security services to itself and to the system
it is embedded in. A SoC can use a TPM to enhance the security
functionality it offers. An application processor can send commands to the
TPM requesting certain operations to be completed securely. It is
important to note that the TPM only uses its resources to provide
functionality to itself or others. The TPM is independent of the system
it is embedded in, both hardware and software wise, and provides dedicated circuitry to protect against physical attacks~\cite{tpm}.

\begin{figure}[htb]
  \centering
  \includegraphics[width=0.2\textwidth]{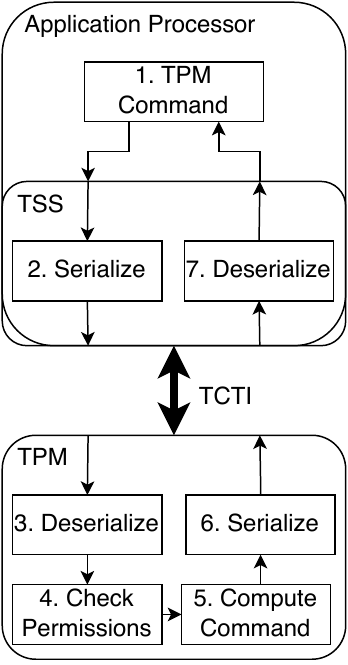}
  \caption{Function flowgraph executed by the TPM when a valid command is received.}
  \label{fig:flow}
\end{figure}

Through the TPM Command Transmission Interface (TCTI) layer, a client is able to interface with a TPM using the commands provided in the TPM Software Stack (TSS). Physically, the TCTI
layer is a bus that connects the application processor to the TPM. This is generally achieved using SPI. Figure~\ref{fig:flow} shows a callgraph of the chain of functions executed when a command is received.
A user will issue a command to the TPM using the TSS. Within this stack, the user's command will be serialized
and sent to the TPM through the TCTI. The TPM, upon receiving the command, will deserialize it and check if the
caller has sufficient privileges to execute this command. If the command passes the check, then the TPM will
execute the command and return the result of the command back to the application processor. The TPM serializes
the results and sends them through TCTI to the application processor which will deserialize it upon reception
using the TSS. Finally, the TSS returns the result of the command to the user.

TPMs can be implemented in different formats~\cite{TCG2016}. Hardware TPMs, besides strengthening the software Trusted Computing Base (TCB) of a system also protect against physical attacks~\cite{TCG2016,MSrecommendations}. Software
TPMs, as is the studied case in this paper, are used for
development and prototyping, since they have a faster implementation-debug cycle than a hardware TPM and also do not have to wear out the non-volatile memory which has a limited lifecycle due to the number of writes~\cite{TCG2016}. 
Moreover, TPMs can also be used in virtualization scenarios where an hypervisor offers the services of a TPM through
a virtualization layer~\cite{TCG2016,MSrecommendations}. This can be achived through multiple Software TPMs running
in the hypervisor for each virtual machine, or the hypervisor uses a hardware TPM in a multi-programmed manner,
context switching part of the state of the TPM between each active virtual machine.

\begin{figure}[htb]
  \centering
  \includegraphics[width=0.25\textwidth]{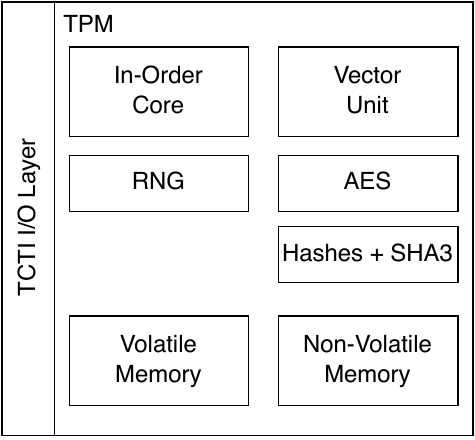}
  \caption{QR TPM Architecture.}
  \label{fig:qr-arch}
\end{figure}

\begin{figure}[htb]
  \centering
  \includegraphics[width=0.30\textwidth]{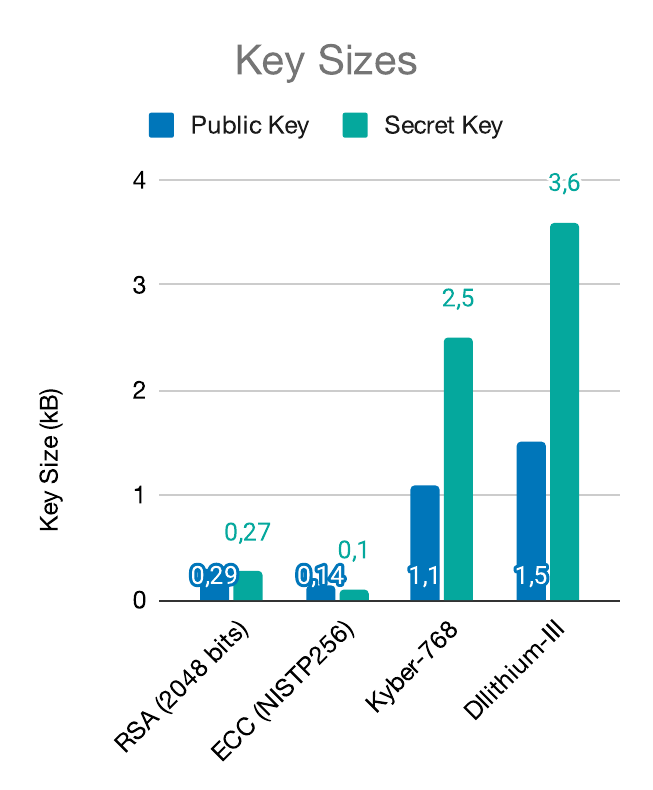}
  \caption{Key size comparison between RSA 2048 bits, ECC NISTP256, Kyber-768, and Dilithium III.}
  \label{fig:key-sizes}
\end{figure}

The general implementation of a TPM features a small in-order CPU, a small on-chip memory, and a cryptographic co-processor to accelerate cryptographic tasks. Number-Theoretic TPMs
have a co-processor that specializes in big-number algebra due to the algebraic requirements of RSA and ECC. QR TPMs, using the soon-to-be standardized Kyber and Dilithium algorithms,
will start moving into co-processors that specialize in lattice-based algebra. Lattice
operations are over a polynomial ring with $n$ dimensions modulus a $q$, wherein $q$ is a prime number.
Algebraic operations over the ring require adding, subtracting, multiplying, and dividing polynomials where all operations are modulus $q$. Therefore, QR TPMs will feature a specialized vector engine in its cryptographic co-processor to handle the ring algebra. Moreover, to improve performance and reduce the number of operations in particular algebraic operations, the co-processor will first convert the polynomials to the NTT domain. Previous work in this area has shown that a QR TPM will possess an architecture that is similar to Figure~\ref{fig:qr-arch} with larger buffers. The median for the extra memory requirements is one
order of magnitude, due to the larger key sizes present in lattice-based cryptography~\cite{fiolhais2020software,ducas19,bos18}. Figure~\ref{fig:key-sizes} shows a public and secret key size
comparison between RSA 2048 bits, ECC NISTP256, Kyber-768, and Dilithium III. However,
the security strength provided by the QR algorithms offsets the increased memory cost. Note that, the specialized accelerator for lattice-based cryptography will vary depending on the order of the polynomial ring and the modulus $q$ supported. As a general rule of thumb, cryptographers increase the order of the polynomial and the modulus $q$ to provide stronger security~\cite{bos18,ducas19}.

\section{TPM Emulator and ROT Implementation}

To emulate the TPM, we use a fork of the IBM Software TPM (SW-TPM) and
TSS~\cite{ibm-tss, ibm-tpm} that supports some QR algorithms in~\cite{sw-tpm, sw-tss}.
The SW-TPM emulates the
TCTI layer through a TCP socket. A command
sends its data to the socket for the TPM to process. Due to the nature of the emulation, the SW-TPM is
also referred to as the TPM server. The SW-TPM attempts to closely emulate the memory limitations
found in a real TPM. There is no reliance on dynamic memory management. The SW-TPM
manages its memory by using buffers allocated either in the stack, or the \texttt{.bss} and
\texttt{.data} program segments. The main reason behind the choice of this fork is because it already
contained the infrastructure to support QR algorithms, as it was used to evaluate the architectural
implementation of QR algorithms such as Kyber~\cite{bos18}, Dilithium~\cite{ducas19},
NTTRU~\cite{nttru}, and L-DAA~\cite{ldaa}.  

The TPM was also extended to add support for a ROT~\cite{Branco21} protocol to evaluate its
efficiency in a TPM scenario. The ROT protocol is
implemented in C++ and Assembly (for certain optimizations)~\cite{roted}. However, the SW-TPM
and its TSS are written in C~\cite{sw-tpm, sw-tss}. C and C++ programs can be linked
together if the C++ program exposes a C Application Binary Interface (ABI). Therefore, we have
modified the ROT protocol such that
it can expose functions for each of the required messages in the protocol. Moreover, we disabled
all optimizations found in the ROT implementation, namely vector instructions for ARM, such that
its computational model would fit the TPM. With these modifications, we compiled a static library
that could be linked with the SW-TPM. Finally, to add support for the ROT algorithm, four new
TPM commands were added for each message passed in the protocol. The \texttt{CC\_ROT\_MSG1} command
computes and transfers the receiver's first message. The \texttt{CC\_ROT\_MSG2} command computes and
transfers the sender's first message. The \texttt{CC\_ROT\_MSG3} command computes and transfers the
receiver's second, and final, message. Lastly, the \texttt{CC\_ROT\_MSG4} command computes the
sender's challenges. Furthermore, similar modifications were employed to the TSS, in the
form of new C binaries and modifications to the TSS library, in order to create new commands to
call the desired ROT functionality in the SW-TPM.

The addition of this new algorithm to the SW-TPM and the TSS totaled 880 lines and 3111 lines
modified, respectively. The implementation of ROT in the SW-TPM and the TSS shows that the usage
of a SW-TPM to prototype features that future TPM specifications may provide requires a small effort. 

\section{Experimental Results}
\label{sec:experimental}

In the ARM core, the compiler used was GCC 8.3.0 and in RISC-V core the compiler used was GCC 11.2.0, both with the \texttt{-O0} optimization flag for
the TPM and the TSS. Even though the ARM cortex possesses a vector unit in the NEON unit, no vector instructions were used either explicitly by the TPM server, the TSS commands, or the compiler. This choice was made because the current TPM 2.0 architecture does
not contain a vector unit. To avoid cache misses due to core migration, the TPM server and the TSS commands were executed
in different cores where each process was pinned to the same core. The frequency was fixed at 900 MHz in the ARM core and at 1.2 GHz in the RISC-V core such that dynamic frequency
scaling would not taint the performance results. 

\begin{figure}
\centering
     \begin{subfigure}[b]{0.3\textwidth}
         \centering
         \includegraphics[width=\textwidth]{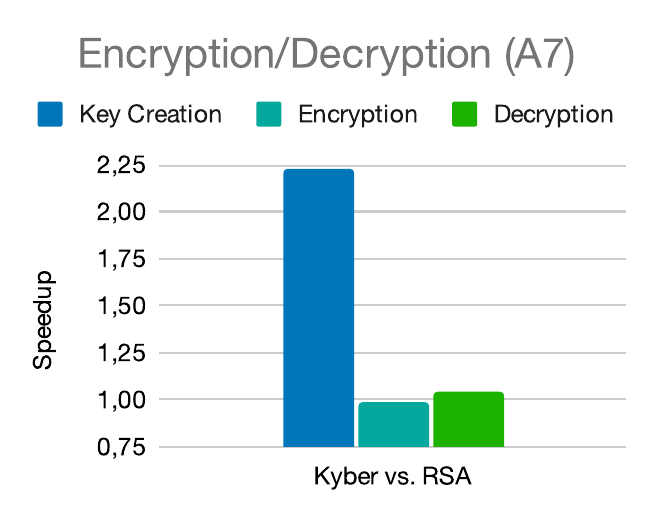}
     \end{subfigure}
     \\
     \begin{subfigure}[b]{0.3\textwidth}
         \centering
         \includegraphics[width=\textwidth]{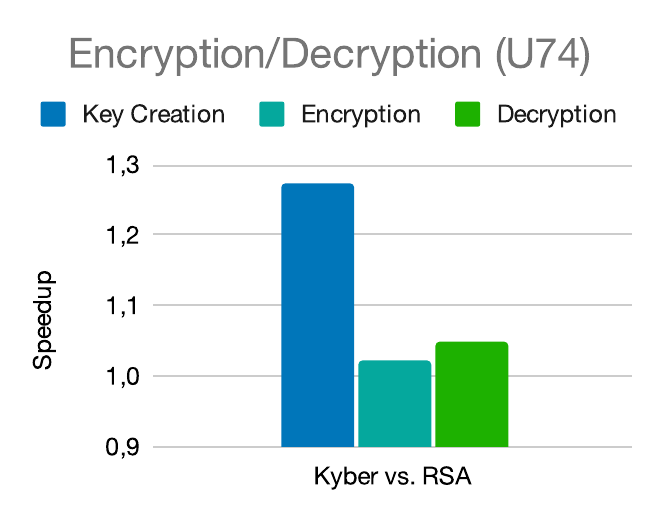}
     \end{subfigure}
     \caption{Encryption and Decryption Speedups between RSA and Kyber for the
     ARM Cortex-A7 and the RISC-V U74.}
    \label{fig:speedup-encryption}
\end{figure}

\begin{figure}
\centering
     \begin{subfigure}[b]{0.3\textwidth}
         \centering
         \includegraphics[width=\textwidth]{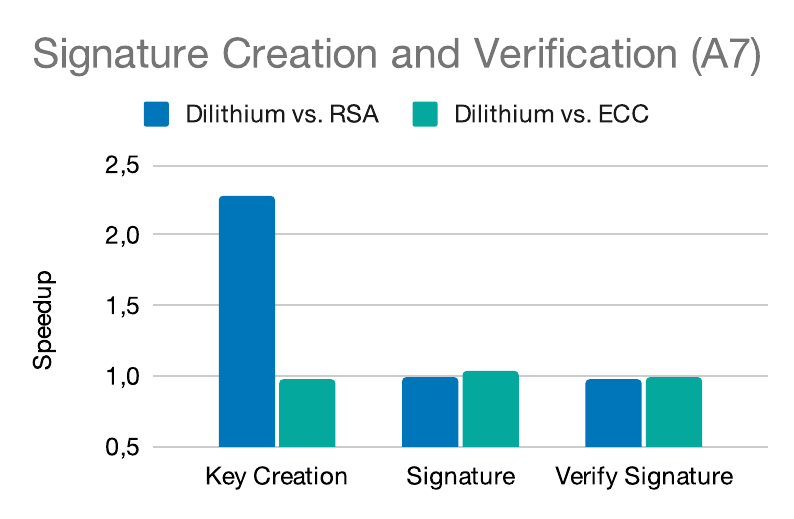}
     \end{subfigure}
     \\
     \begin{subfigure}[b]{0.3\textwidth}
         \centering
         \includegraphics[width=\textwidth]{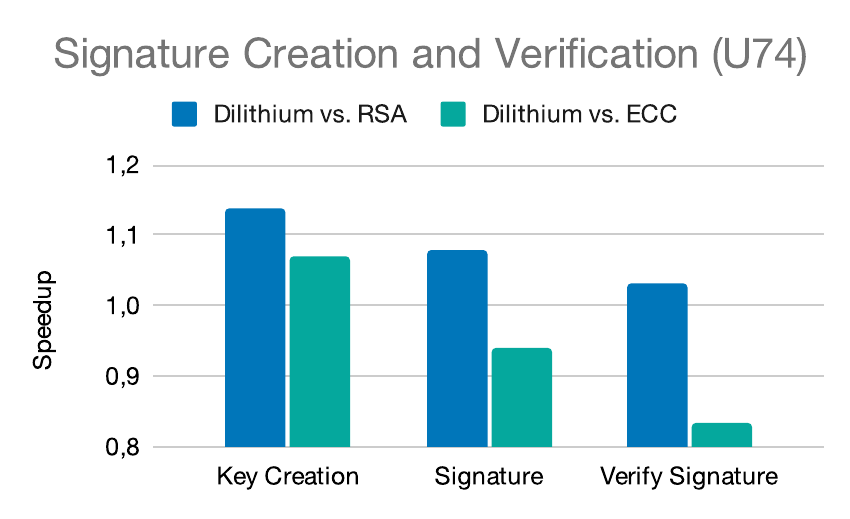}
     \end{subfigure}
     \caption{Signature Creation and Verification Speedups between RSA, ECC, 
     and Dilithium for the ARM Cortex-A7 and the RISC-V U74.}
    \label{fig:speedup-signature}
\end{figure}

The experiment methodology used herein is the following.
The security parameters used for each algorithm are:
2048 bits for RSA, the NISTP256 curve for ECC, the 768 mode in Kyber
($n = 256$ and $q = 3329$)~\cite{bos18}, the level III mode
in Dilithium ($n = 256$ and $q = 8380417$)~\cite{ducas19}, and the default security parameters in
ROT ($n = 512$ and $q = 13313$)~\cite{Branco21}.
The ASCII string "My super secret. Please don’t share.\textbackslash n" is used for encryption and signature; signed messages use the SHA3-256 hash; and all keys are created as non-primary with the fixed TPM and parent
properties. All the measured times result from taking the median over one hundred runs running on each
of the previously described processors. The used TSS commands are: \texttt{CC\_Create} for key creation;
\texttt{CC\_Sign} for data signature; \texttt{CC\_VerifySignature} for signature verification; 
\texttt{CC\_\{KYBER, RSA\}\_Encrypt} and \texttt{CC\_\{KYBER, RSA\}\_Decrypt} for Kyber and RSA encryption
and decryption; \texttt{CC\_KYBER\_Enc} and \texttt{CC\_KYBER\_Dec} for Kyber 
encapsulation and decapsulation; and \texttt{CC\_\{ROT\_MSG1, ROT\_MSG2,
ROT\_MSG3, ROT\_MSG4\}} for the ROT operations. The performance results for each QR and number-theoretic
algorithm can be found in Table~\ref{tab:arm-results} for the ARM processor, and in 
Table~\ref{tab:riscv-results} for the RISC-V processor. Furthermore, Figures~\ref{fig:speedup-encryption}
and~\ref{fig:speedup-signature} shows the speedups for each core between QR and number-theoretic algorithms.
Note that, in the ROT protocol the time for the receiver
is the result of the addition of the time for \texttt{MSG1} and \texttt{MSG3}, and
the time for the sender is the result of the addition of the time for \texttt{MSG2}
and \texttt{MSG4}.

\begin{table}[htb]
\caption{Execution time (s) for number-theoretic and post-quantum lattice-based algorithms in a TPM 2.0 running
in an ARM Cortex-A7 processor.}
\label{tab:arm-results}
\resizebox{\columnwidth}{!}{%
\begin{tabular}{|c|c|c|c|c|c|}
\hline
\textbf{}        & \textbf{RSA} & \textbf{ECC } & \textbf{Kyber} & \textbf{Dilithium} & \textbf{ROT} \\ 
\textbf{}        & \textbf{2048 bits} & \textbf{NISTP256} & \textbf{768} & \textbf{III} & \textbf{} \\ \hline
\textbf{Key Creation}     & 3,54                     & 1,59                    & 1,59              & 1,55                   & -            \\ \hline
\textbf{Signature}        & 1,59                     & 1,54                    & -                 & 1,59                   & -            \\ \hline
\textbf{Verify Signature} & 1,52                     & 1,55                    & -                 & 1,55                   & -            \\ \hline
\textbf{Encryption}       & 1,54                     & -                       & 1,56              & -                      & -            \\ \hline
\textbf{Decryption}       & 1,61                     & -                       & 1,55              & -                      & -            \\ \hline
\textbf{ROT Receiver}       & -                        & -                       & -                 & -                      & 3,16         \\ \hline
\textbf{ROT Sender}     & -                        & -                       & -                 & -                      & 3,12         \\ \hline
\end{tabular}%
}
\end{table}

\begin{table}[htb]
\caption{Execution time (s) for number-theoretic and post-quantum lattice-based algorithms in a TPM 2.0 running
in a RISC-V U74 processor.}
\label{tab:riscv-results}
\resizebox{\columnwidth}{!}{%
\begin{tabular}{|c|c|c|c|c|c|}
\hline
\textbf{}        & \textbf{RSA} & \textbf{ECC} & \textbf{Kyber} & \textbf{Dilithium} & \textbf{ROT} \\
\textbf{}        & \textbf{2048 bits} & \textbf{NISTP256} & \textbf{768} & \textbf{III} & \textbf{ROT} \\ \hline
\textbf{Key Creation}     & 4,00                     & 3,28                    & 3,14              & 3,51                   & -            \\ \hline
\textbf{Signature}        & 3,22                     & 3,17                    & -                 & 2,98                   & -            \\ \hline
\textbf{Verify Signature} & 3,23                     & 3,75                    & -                 & 3,13                   & -            \\ \hline
\textbf{Encryption}       & 3,12                     & -                       & 3,05              & -                      & -            \\ \hline
\textbf{Decryption}       & 3,29                     & -                       & 3,14              & -                      & -            \\ \hline
\textbf{ROT Receiver}       & -                        & -                       & -                 & -                      & 6,51         \\ \hline
\textbf{ROT Sender}     & -                        & -                       & -                 & -                      & 7,1          \\ \hline
\end{tabular}%
}
\end{table}

In the ARM core, the QR algorithms Kyber and Dilithium fully replace the functionality provided by
RSA and ECC. Regarding performance, Kyber and Dilithium are faster at creating keys than RSA, due to the key size and secure random sampling required in RSA. However, they meet the same
performance level as ECC. For digital signatures, both in signature creation and verification,
Dilithium is on par with both RSA and ECC. In data encryption and decryption, Kyber is also on par
with RSA. With respect to ROT, since the TPM does not yet offer commands or requires an  implementation there is no one-to-one comparison with a number-theoretic counterpart. However, it has been previously shown that QR OTs and ROTs significantly outperform number-theoretic OTs
and ROTs~\cite{Branco21}. Therefore, only the QR variants are analyzed. Still, the performance results for the ROT commands both for a sender and a receiver are approximately double that of Kyber and Dilithium even though they possess more complex lattice arithmetic. The ROT results show that it would be possible for a QR TPM to support ROTs and OTs with minimal architectural efforts and quite decent performance.

In the RISC-V U74 core, there are some improvements in the QR algorithms. Kyber and Dilithium create their keys faster than RSA but on par with ECC. However, in signature operations, Dilithium shows a  speedup of at least 6\% when compared with ECC and RSA. The same is true for encryption operations. Kyber outperforms RSA by at least 3\%. For ROT operations, the same conclusions can be drawn from the ARM experiments, despite ROT having more complex
operations than Kyber and Dilithium, it still provides quite reasonable performance.

\section{Conclusions}
\label{sec:conclusions}
This paper showed how to extend the current Number-Theoretic base TPMs to be QR. By substituting the RSA and ECC algorithms by the Kyber, for key encapsulation and data encryption/decryption, and the Dilithium primitives, for digital signature, quantum resistance is ensured by the hardness of mathematical problems related to lattices. Moreover, the TPM was also extended, for the first time, to add support for a ROT protocol, useful for example to  Multi-Party Computation, which is based on the same type of mathematical problems. Not only were the computational and memory requirements for this TPM extension analysed, but it was experimentally evaluated by implementations on ARM and RISC-V low-power processors. It is shown that the security strength provided by the QR algorithms increased memory requirements, while maintaining or even decreasing the execution time of the QR algorithms. The ROT protocol
was implemented without significant changes to the architectural model of the TPM and it possesses decent
performance. Finally, we show that the usage of programmable microprocessors in future TPMs would allow a vendor
to add new algorithms remotely to older hardware TPMs, in order to improve its security features, with minimal
changes to the architectural model. This paper paves the way for the design of QR TPMS based on low-power
programmable devices.


\bibliographystyle{ACM-Reference-Format}
\bibliography{main}

\end{document}